\documentclass[aps,prl,twocolumn,superscriptaddress,showpacs]{revtex4}

\usepackage{graphicx}
\usepackage{dcolumn}
\usepackage{bm}

\begin{document}


\title{Double-beta decay $Q$ values of $^{130}$Te, $^{128}$Te, and $^{120}$Te}



\author{N.~D.~Scielzo}
\affiliation{Physical Sciences Directorate, Lawrence Livermore National Laboratory, Livermore, California 94550, USA}

\author{S.~Caldwell}
\affiliation{Physics Division, Argonne National Laboratory, Argonne, Illinois 60439, USA}
\affiliation{Department of Physics, University of Chicago, Chicago, Illinois 60637, USA}

\author{G.~Savard}
\affiliation{Physics Division, Argonne National Laboratory, Argonne, Illinois 60439, USA}
\affiliation{Department of Physics, University of Chicago, Chicago, Illinois 60637, USA}


\author{J.~A.~Clark}
\affiliation{Physics Division, Argonne National Laboratory, Argonne, Illinois 60439, USA}


\author{C.~M.~Deibel}
\affiliation{Physics Division, Argonne National Laboratory, Argonne, Illinois 60439, USA}
\affiliation{Joint Institute of Nuclear Astrophysics, Michigan State University, East Lansing, MI 48824, USA}

\author{J.~Fallis}
\affiliation{Physics Division, Argonne National Laboratory, Argonne, Illinois 60439, USA}
\affiliation{Department of Physics and Astronomy, University of Manitoba, Winnipeg, Manitoba R3T 2N2, Canada}

\author{S.~Gulick}
\affiliation{Department of Physics, McGill University, Montr$\acute{e}$al, Qu$\acute{e}$bec H3A 2T8, Canada}

\author{D.~Lascar}
\affiliation{Physics Division, Argonne National Laboratory, Argonne, Illinois 60439, USA}
\affiliation{Department of Physics and Astronomy, Northwestern University, Evanston, Illinois 60208, USA}


\author{A.~F.~Levand}
\affiliation{Physics Division, Argonne National Laboratory, Argonne, Illinois 60439, USA}

\author{G.~Li}
\affiliation{Physics Division, Argonne National Laboratory, Argonne, Illinois 60439, USA}
\affiliation{Department of Physics, McGill University, Montr$\acute{e}$al, Qu$\acute{e}$bec H3A 2T8, Canada}

\author{J.~Mintz}
\affiliation{Department of Nuclear Engineering, University of California, Berkeley, California 94720, USA}

\author{E.~B.~Norman}
\affiliation{Physical Sciences Directorate, Lawrence Livermore National Laboratory, Livermore, California 94550, USA}
\affiliation{Department of Nuclear Engineering, University of California, Berkeley, California 94720, USA}


\author{K.~S.~Sharma}
\affiliation{Department of Physics and Astronomy, University of Manitoba, Winnipeg, Manitoba R3T 2N2, Canada}

\author{M.~Sternberg}
\affiliation{Physics Division, Argonne National Laboratory, Argonne, Illinois 60439, USA}
\affiliation{Department of Physics, University of Chicago, Chicago, Illinois 60637, USA}

\author{T.~Sun}
\affiliation{Physics Division, Argonne National Laboratory, Argonne, Illinois 60439, USA}

\author{J.~Van Schelt}
\affiliation{Physics Division, Argonne National Laboratory, Argonne, Illinois 60439, USA}
\affiliation{Department of Physics, University of Chicago, Chicago, Illinois 60637, USA}


\date{\today}

\begin{abstract}

The double-beta decay $Q$ values of $^{130}$Te, $^{128}$Te, and $^{120}$Te have been determined from parent-daughter mass differences 
measured with the Canadian Penning Trap mass spectrometer.  The $^{132}$Xe--$^{129}$Xe mass difference, which is precisely known, was
also determined to confirm the accuracy of these results.  The $^{130}$Te $Q$ value was found to be $2527.01\pm0.32$ keV which is 
3.3 keV lower than the 2003 Atomic Mass Evaluation recommended value, but in agreement with the most precise previous measurement.  The 
uncertainty has been reduced by a factor of 6 and is now significantly smaller than the resolution achieved or foreseen in experimental 
searches for neutrinoless double-beta decay.  The $^{128}$Te and $^{120}$Te $Q$ values were found to be $865.87\pm1.31$ keV and 
$1714.81\pm1.25$ keV, respectively.  For $^{120}$Te, this reduction in uncertainty of nearly a factor of 8 opens up the 
possibility of using this isotope for sensitive searches for neutrinoless double-electron capture and electron capture with $\beta^{+}$ 
emission.

\end{abstract}

\pacs{21.10.Dr,23.40.-s,27.60.+j,37.10.Ty}

\maketitle




A definitive observation of neutrinoless double-beta ($0\nu\beta\beta$) decay would have many profound implications such as revealing the Majorana 
nature of the neutrino, constraining the neutrino mass hierarchy and scale, and providing a mechanism for the violation of lepton number 
conservation (see Refs. \cite{el00,el04,av08} for recent reviews).
Currently the best $0\nu\beta\beta$-decay half-life limits come from searches for the characteristic energy peak at the $Q$ values 
of $^{130}$Te \cite{ar08} or $^{76}$Ge \cite{ba99,aa02,aa04}.  The $^{76}$Ge $Q$ value has been determined several times \cite{do01,su07,ra08} 
with a precision significantly better than the 1 keV resolution (1 $\sigma$) of the enriched HPGe detectors used in these experiments.  The 
$^{130}$Te $Q$ value, however, rests on much shakier ground.  The 2003 Atomic Mass Evaluation (AME2003) recommended value \cite{au03} has a 
1.99 keV uncertainty and is dominated by 
a single measurement \cite{dy90}.  This uncertainty is comparable to the 2--3 keV resolution (1 $\sigma$) of the TeO$_{2}$ bolometric 
detectors used in the Cuoricino $\beta\beta$-decay experiment \cite{ar08}.  The ton-scale CUORE TeO$_{2}$ bolometer array \cite{ar04} and 
the COBRA CdTeZn semiconductor array \cite{zu01} anticipate 1 $\sigma$ experimental resolutions (at 2.5 MeV) of approximately 2 keV and 10 keV, 
respectively.

Experiments typically use natural tellurium and therefore are also sensitive to the signatures of 
$^{128}$Te $\beta\beta$-decay as well as $^{120}$Te double-electron capture ($\varepsilon\varepsilon$) and electron capture with $\beta^{+}$ 
($\varepsilon\beta^{+}$) decay processes.  The recommended $^{128}$Te $Q$ value is also dominated by a single measurement \cite{dy90}.  
The situation for $^{120}$Te is even worse: the AME2003 recommended $Q$ value has an uncertainty of 10 keV.  Thus, a neutrinoless-decay 
signal could be obscured by or confused with a background line.

To eliminate these concerns, we have determined the $\beta\beta$-decay $Q$ values of $^{130}$Te, $^{128}$Te, and $^{120}$Te by measuring the 
parent-daughter mass differences using the Canadian Penning Trap (CPT) mass spectrometer \cite{sa97}.  High-precision mass spectrometry 
was performed by measuring the cyclotron frequencies $\omega=\frac{qB}{M}$ of ions of mass $M$ and charge $q$ in the homogeneous magnetic field 
$B$ of a Penning trap.  The cyclotron frequencies of the singly-charged parent ($\omega_{1}$) and daughter ($\omega_{2}$) ions determine the 
$Q$ values from the relation:
\begin{equation}
\label{eq:Q-value}
Q=m_{1}-m_{2}=\big(m_{2}-m_{e}\big)\bigg(\frac{\omega_{2}}{\omega_{1}}-1\bigg)
\end{equation}
where $m_{1}$, $m_{2}$, and $m_{e}$ are the masses of the neutral parent atom, neutral daughter atom, and electron, respectively.  
The valence electron binding energies were 7--12 eV and can be neglected here.  The ratios $\frac{\omega_{2}}{\omega_{1}}$ were measured 
to fractional precisions approaching $10^{-9}$ using the techniques described below.  The CPT mass spectrometer has previously 
been used to measure the masses of short-lived isotopes to fractional precisions of $10^{-7}$ to $10^{-9}$ to determine (single) 
$\beta$-decay $Q$ values \cite{sa04,sa05} and proton separation energies \cite{cl04,cl07,fa08}.  


The CPT mass spectrometer has been described in detail in several publications \cite{sa97,sa01,cl03,cl04}.  
The ion preparation and measurement technique is briefly presented here with details provided for the aspects of the experiment 
developed for this work.

Singly-charged ions of Te and Sn were produced by laser ablation of solid targets using a Q-switched Nd:YAG laser
to deliver up to 0.5 mJ of energy in 5-ns pulses at a 20-Hz repetition rate.  Three targets were mounted on a movable frame:  
two pressed and lacquer-bound tellurium metal powder targets (one with $^{\mathrm{nat}}$Te for $^{130}$Te and $^{128}$Te 
measurements and one enriched to 37.4\% in $^{120}$Te) and one $^{\mathrm{nat}}$Sn foil.  The samples were mechanically 
rastered to maintain consistent ion output over extended periods of time.  For the $^{130}$Xe and $^{128}$Xe measurements, 
$^{\mathrm{nat}}$Xe gas was injected directly into a radiofrequency quadrupole (RFQ) ion guide \cite{sa01} and ionized by 
a nearby ion gauge.  

Ions from either the RFQ ion guide or the solid targets then entered a 1.5-kV electrostatic beamline.  
A voltage pulse applied to one of the beamline electrodes was timed to allow only ions with the 
desired mass number to reach the purification trap.  This trap, a cylindrical Penning trap filled with helium 
buffer gas, accumulated and thermalized the ion bunches.  An RF field applied at the appropriate cyclotron frequency 
centered the ions of interest while the contaminant ions were driven out of the trap \cite{sa91}.  The purified ion bunches were then 
transported to a linear RFQ ion trap \cite{cl03} filled with helium buffer gas.  This trap 
accumulated and cooled the captured ion bunches and ensured conditions were identical for all isotopes injected into the CPT.

Ions loaded in the CPT were confined radially by a constant magnetic field ($B=5.9$ T) and axially by a quadrupole 
electrostatic potential, resulting in reduced-cyclotron, magnetron, and axial eigenmotions \cite{br86}.  An ``evaporation pulse'' 
adiabatically reduced the depth of the electrostatic trapping potential 
for about 10 ms to expel $>$90\% of the ions.  Only the coldest ions, which occupied the smallest volume 
at the trap center, remained.  Next, dipole RF fields were applied at reduced-cyclotron frequencies to mass-selectively 
drive any remaining unwanted ion species from the trap.

The cyclotron frequency of ions in the CPT was determined using a time-of-flight (TOF) measurement 
technique \cite{gr80}.  The ions were first excited to a magnetron orbital radius using a dipole RF field.
A quadrupole RF field at a frequency $\omega_{ex}$ near $\omega_{c}$ was then applied.  On resonance, the RF field fully converted magnetron 
motion to reduced cyclotron motion.  The ions were subsequently ejected from the trap and allowed to drift toward a microchannel plate 
detector (MCP) located in a region of lower $B$ field.  As ions traversed the $B$-field gradient, the kinetic 
energy of the cyclotron motion was converted to linear kinetic energy.  Near the MCP, ions were accelerated to 
2.4 keV for detection and the TOF recorded using a multi-channel scaler.  
The ion TOF was smallest when the conversion to reduced cyclotron motion was most effective.  The resonance shape was determined from 
repeated measurements of the TOF spectrum at a series of equally-spaced $\omega_{ex}$ values.

\begin{figure}
\includegraphics[width=0.50\textwidth]{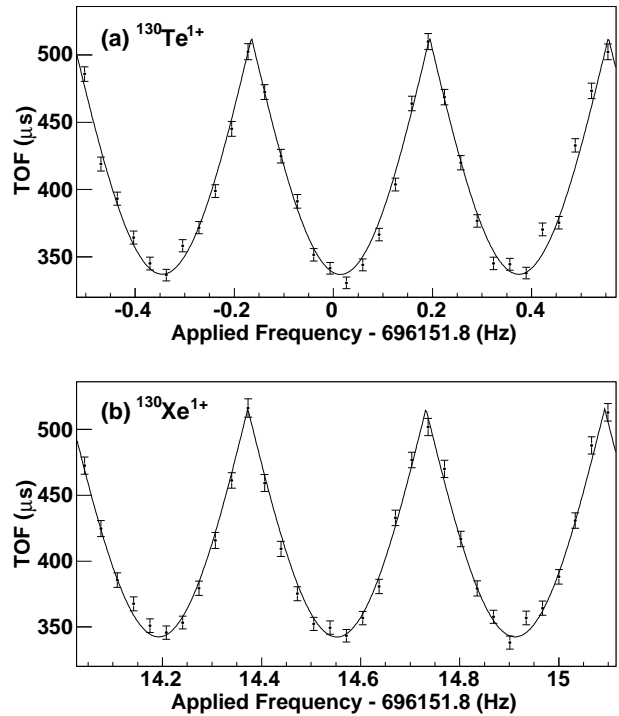}
\caption{\label{fig:Te130_TOF}  Time-of-flight spectra for (a) $^{130}$Te$^{1+}$ and (b) $^{130}$Xe$^{1+}$ ions using two-pulse Ramsey 
excitations over a total time of 3000 ms.  Only events with $\leq$5 detected ions were included in these spectra.  A fixed TOF offset, 
common to all the data, was not measured and has no impact on the determination of cyclotron frequencies.  The smooth curves 
are the fits to the data.}
\end{figure}

The two-pulse Ramsey method of separated oscillatory fields \cite{ra90} was applied for the $\omega_{ex}$ quadrupole 
RF-field excitation.  The resulting TOF resonance pattern is shown in Fig. \ref{fig:Te130_TOF}.  The use of two pulses 
resulted in the narrowest linewidths (for a given excitation time) which increased the precision of the measurements \cite{su07,ge07,kr07}.  
The central TOF minimum was identified from an initial scan using only one excitation pulse which yielded a single, dominant TOF minimum.  
Recently, other heavy-ion mass measurements have also applied this Ramsey method \cite{ge07,ge07_2,ra08}.
%


The ratio of measured cyclotron frequencies and the resulting $Q$ values determined by Eq. \ref{eq:Q-value} are listed in
Table \ref{ta:Results}.  The experiments were performed under two sets of measurement conditions.
The $^{130}$Te--$^{130}$Xe measurements 
and initial $^{120}$Te--$^{120}$Sn measurements were performed with a probe time of 3000 ms (two 300-ms excitation pulses separated 
in time by 2400 ms).  The TOF spectra for $^{130}$Te and $^{130}$Xe measurements with $\leq$5 detected ions per shot are shown in 
Fig. \ref{fig:Te130_TOF}.

A second set of measurements ($^{128}$Te--$^{128}$Xe and $^{120}$Te--$^{120}$Sn) used a 1000-ms sequence of two 
100-ms pulses separated by 800 ms.  The use of a shorter probe time was necessitated by a small distortion in the electric field of 
the CPT that broadened the resonance linewidth.  This distortion occurred after the data collection with 3000-ms probe times and
did not affect this earlier data.

Under both sets of conditions, the accuracy of the technique was verified by measuring the $^{132}$Xe--$^{129}$Xe mass difference.
The mass difference was found to be $2793899.12\pm0.30$ keV and $2793900.22\pm0.97$ keV (including systematic 
uncertainties discussed below) for the 3000-ms and 1000-ms measurements, respectively.  These results agree with the ultra-precise 
value $2793899.180\pm0.070$ keV from Ref. \cite{re09} using $931494.028\pm0.023$ keV/c$^{2}$ per u \cite{mo08}.  The agreement 
demonstrates that systematic uncertainties are under control and all the $\beta\beta$-decay $Q$ values determined in this work are 
reliable.

\begin{table}
\caption{\label{ta:Results} Summary of measured cyclotron frequency ratios $\omega_{2}$/$\omega_{1}$ and the resulting mass 
differences $\Delta$$M$ measured using two-pulse Ramsey method times of 3000 ms and 1000 ms.  The measured 
$^{132}$Xe--$^{129}$Xe mass differences agreed with the value 2793899.180(70) keV from Ref. \cite{re09} using 931494.028(23) 
keV/c$^{2}$ per u \cite{mo08}.  The $^{130}$Xe--$^{129}$Xe mass difference was used to determine the absolute mass of $^{130}$Te
and $^{130}$Xe.  The uncertainties include both statistics and systematics.}
\begin{ruledtabular}
\begin{tabular}{lccr}
Isotopes & Time & $\omega_{2}$/$\omega_{1}$ & $\Delta$$M$ (keV) \\
\hline
$^{130}$Te--$^{130}$Xe & 3000 ms & 1.0000208837(26) & 2527.01(32) \\
$^{120}$Te--$^{120}$Sn & 3000 ms & 1.0000153639(142) & 1715.96(159) \\
$^{132}$Xe--$^{129}$Xe & 3000 ms & 1.0232682365(25) & 2793899.12(30) \\
$^{130}$Xe--$^{129}$Xe & 3000 ms & 1.0077478330(26) &  930309.60(32) \\
$^{120}$Te--$^{120}$Sn & 1000 ms & 1.0000153436(141) & 1713.69(157) \\
$^{128}$Te--$^{128}$Xe & 1000 ms & 1.0000072676(110) & 865.87(131) \\
$^{132}$Xe--$^{129}$Xe & 1000 ms & 1.0232682457(81) & 2793900.22(97) \\
\end{tabular}
\end{ruledtabular}
\end{table}

Most systematic effects are expected to cancel in the frequency ratios because the measurements were performed 
under identical experimental conditions.  The results of the $^{132}$Xe--$^{129}$Xe measurements confirmed that mass-dependent 
systematic effects were $\lesssim0.3$ keV/u; this is consistent with previous CPT studies \cite{va02}.  Mass-dependent systematic 
effects were therefore negligible because the $\beta\beta$-decay mass differences ($\lesssim0.003$ u) were $\times10^{3}$ smaller. 

The time and ion-number dependence of the results were investigated in detail.  Parent and daughter measurements were interleaved 
in time to minimize effects such as $B$-field drift.  Over the course of the experiment, 
the measured cyclotron frequencies showed no evidence of drifts.  Upper limits on the drifts were 2.5 ppb/day for the 
$^{130}$Te--$^{130}$Xe pair and 7 ppb/day for the other isotopes.  These limits are consistent with previous $B$-field stability measurements
\cite{fa08}. Uncertainties of 0.08 keV and 0.2 keV for the 3000-ms and 1000-ms measurements, respectively, 
result from the timing of the data collection.   

\begin{figure}
\includegraphics[width=0.5\textwidth]{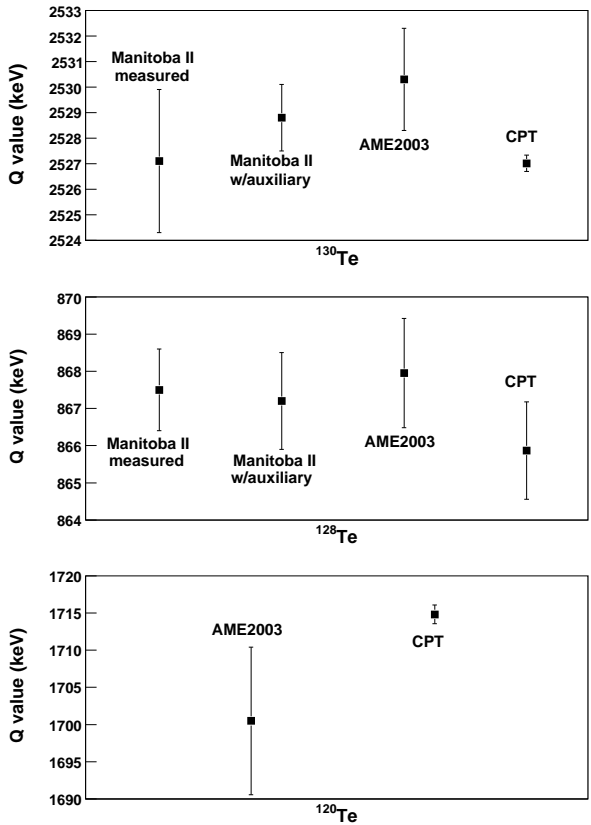}
\caption{\label{fig:Te130_Qvalues} $Q$ values for $^{130}$Te, $^{128}$Te, and $^{120}$Te determined from the most recent measurements 
and evaluations.  The values shown are the result of the mass splitting measurement using the Manitoba II deflection-type mass 
spectrometer (Manitoba II measured) \cite{dy90}, the Manitoba II result supplemented with auxiliary mass data (Manitoba II w/auxiliary) 
\cite{dy90}, the 2003 Atomic Mass Evaluation (AME2003) \cite{au03}, and this work (CPT).}
\end{figure}

The ion detection rate was intentionally kept low (typically $<$10 ions detected per bunch) to limit any frequency dependence 
on trapped ion population.  Any shift caused by the added space charge from multiple ions trapped simultaneously was expected to be 
identical for parent and daughter measurements.  The results were sorted event-by-event by ion number and cyclotron frequencies were 
determined from these spectra.  
For the 3000-ms measurements, linear shifts were consistent with zero with a value of $-0.05\pm0.05$ keV ($-0.4\pm0.4$ ppb) per detected ion 
were observed.  For the 1000-ms measurements, resolved linear shifts were observed; for each isotope, the measured slope was consistent with 
$1.5\pm0.2$ keV ($13\pm2$ ppb) per detected ion.  To eliminate these shifts, frequency ratios and $Q$ values (listed in  
Table \ref{ta:Results}) were determined at each ion number for data with $\leq$5 detected ions and the results determined from the weighted 
average.  For parent-daughter measurements performed with identical trapped-ion distributions, these shifts would completely cancel in the 
frequency ratios and therefore $Q$ values.  The ion distributions in these measurements, although similar (all had an average of nearly 2
detected ions per bunch), were not identical.  To account for possible imperfect cancellations of the ion-number shift, we assign 
systematic uncertainties of 0.1 keV and 0.4 keV for 3000-ms and 1000-ms measurements, respectively.  These values are based on the 
consistency of the frequency shifts (at 2 ions detected per bunch) for the measurements.  In each case, the total uncertainty was 
dominated by statistics.  The results of these measurements are discussed below and displayed in Fig. \ref{fig:Te130_Qvalues}.

\emph{$^{130}$Te--$^{130}$Xe}:  The $Q$ value we determine ($2527.01\pm0.32$ keV) agrees with the mass difference 
measured using the Manitoba II deflection-type mass spectrometer \cite{ba71,dy90} (see Fig. \ref{fig:Te130_Qvalues}).  
In \cite{dy90}, an evaluation of the data incorporating auxiliary $^{124,126,128}$Te and $^{128}$Xe measurements increased the
central value by 1.7 keV.  In the AME2003 \cite{au03}, the incorporation of modern auxiliary data further increased 
the recommended $Q$ value to $2530.30\pm1.99$ keV.  Our result is 6$\times$ more precise and 3.3 keV smaller than the AME2003 
recommended value.

Using the $^{129}$Xe mass in Ref. \cite{re09} to set the absolute scale for our 3000-ms measurements, the masses of 
$^{130}$Te and $^{130}$Xe were determined to be 129906222.18$\pm$0.48 $\mu$u and 129903509.32$\pm$0.34 $\mu$u, respectively.  
The 3.3 keV shift in the $Q$ value is the result of the $^{130}$Te mass being 
2.22 $\mu$u smaller than the AME2003 value of $129906224.399\pm2.067$ $\mu$u and the $^{130}$Xe mass being 1.32 $\mu$u larger 
than the value of $129903508.007\pm0.805$ $\mu$u.  Recent $^{129,132}$Xe results \cite{re09} were also $\approx1.5$ $\mu$u larger 
than the AME2003 values. 

This $Q$ value will impact the $0\nu\beta\beta$-decay limit obtained from the Cuoricino experiment by shifting the decay 
energy closer to the $^{60}$Co sum peak background at 2505.7 keV and out of a valley presumably caused by a statistical 
fluctuation \cite{ar08}.  The only isotope in any natural decay chain that emits a $\gamma$-ray within 5 keV of this $Q$ 
value is $^{214}$Bi which may have a weak transition at $2529.7\pm0.8$ keV \cite{ek08}.

\emph{$^{120}$Te--$^{120}$Sn}:  Measurements of the $^{120}$Te $Q$ value were performed with both 3000-ms and 1000-ms timings.  
The results agree (see Table \ref{ta:Results}) and a weighted average yields a $Q$ value of $1714.81\pm1.25$ keV.  This value is 15 keV larger 
than the AME2003 value of $1700.49\pm9.92$ keV \cite{au03} where the uncertainty is dominated by the uncertainty in the $^{120}$Te 
mass.  This value provides guidance for experimental searches for $\varepsilon\varepsilon$ and $\varepsilon\beta^{+}$ decay in $^{120}$Te 
\cite{bl07,ba07,sc08}.  The uncertainty is smaller than the 1 $\sigma$ detector resolution of any fielded or anticipated 
experiment.  

\emph{$^{128}$Te--$^{128}$Xe}:  The AME2003 recommended $Q$ value ($867.95\pm1.47$ keV) has remained nearly unchanged from the Manitoba 
II measurement ($867.5\pm1.1$ keV or $867.2\pm1.3$ keV with auxiliary data) \cite{dy90}.  Our result ($865.87\pm1.31$ keV) 
confirms the accepted value.


In summary, we have measured the $Q$ values for the three naturally-occuring tellurium isotopes for which double-beta decay processes are 
energetically allowed.  We found the $^{130}$Te $Q$ value to be 3.3 keV lower than the 
value recommended in AME2003.  The $Q$-value uncertainty is now significantly smaller than the resolution of any foreseen experiment. The 
$^{128}$Te $Q$ value was also remeasured and found to be consistent with the AME2003 recommended value.

In addition, our $^{120}$Te $Q$-value result reduces the uncertainty by nearly an order of magnitude and will allow 
$\varepsilon\varepsilon$ and $\varepsilon\beta^{+}$ decay searches to be performed with greater sensitivity.
 The $Q$-value energy fortunately lies between the background lines at 1684.0, 1693.1 (single-escape line), and 1729.6 keV
that are common in low-background experiments from the decay of $^{214}$Bi (a $^{222}$Rn daughter).  The sensitivity of the aforementioned 
$0\nu\beta\beta$-decay experiments should be sufficient to place the most stringent limits (achieved with any isotope) for 
$\varepsilon\varepsilon$- and $\varepsilon\beta^{+}$-decay processes.

\begin{acknowledgments}
We thank John Greene for help making the tellurium powder targets.
This work was performed under the auspices of the U.S. Department of Energy by Lawrence Livermore 
National Laboratory under Contract DE-AC52-07NA27344, Argonne National Laboratory under Contract
DE-AC02-06CH11357, and Northwestern University under Contract DE-FG02-98ER41086.  This work was 
supported by NSERC, Canada, under application number 216974.
\end{acknowledgments}

\bibliography{TeQValue.bib}

\end{document}